\documentclass[aps,prb,twocolumn,groupedaddress,maxnames=1]{revtex4-1}
\usepackage{graphicx}
\usepackage{natbib}
\usepackage{longtable}

\begin{document}


\title{Novel superhard sp$^3$-carbon allotropes with odd and even ring topologies}



\author{Daniele Selli$^1$, Igor A. Baburin$^{1,2}$, Roman
  Marto\v{n}\'{a}k$^3$, Stefano Leoni$^{1,}$}
\email{stefano.leoni@chemie.tu-dresden.de} \affiliation{ 
$^1$Technische
  Universit\"at Dresden,Institut f\"ur Physikalische Chemie,01062
  Dresden,Germany\\
  $^2$Max-Planck-Institut f\"ur Chemische Physik fester
  Stoffe, 01187 Dresden, Germany\\
  $^3$Department of Experimental Physics, Comenius University, Mlynsk\'{a}
  Dolina F2, 842 48 Bratislava, Slovakia}

\date{\today}

\begin{abstract}
  Four novel sp$^3$-carbon allotropes with 6, 8 and 16 atoms per primitive
  cell have been derived using a combination of metadynamics simulations
  and topological scan. A novel chiral orthorhombic phase oC16 (C222$_1$) was
  found to be harder than monoclinic M-carbon and shows remarkable
  stability in the high pressure range. A second orthorhombic phase of
  $Cmmm$ symmetry, by $\sim$0.028 eV/atom energetically lower than W-Carbon,
  can be formed from graphite at $\sim$9GPa.  In general, the mechanical
  response under pressure was found to depend on the structure topology,
  which reflects the way rings are formed from an initial graphene layer
  stacking.
\end{abstract}


\maketitle

\section{Introduction}
The quest for novel carbon materials with improved mechanical properties and tailored optical gap is a topic of high priority. Engineering new properties is tightly connected with the ability to predict crystal structures, which remains a crucial issue in both basic solid state research and modern materials science\cite{p2}.

In the effort of anticipating superior materials for catalysis, hydrogen storage and gas segregation, structure prediction stands out for its capacity to efficiently indicate viable technological target compounds. The challenge consists in identifying metastable modifications that can exhibit interesting physical and chemical properties.

Compression of graphite at high pressure and temperatures produces diamond\cite{p1}. Graphite cold compression on the contrary produces a hard and transparent product, different either from cubic or hexagonal diamond\cite{p3,p4,p5,p6,p7,p8,p9}, but not fully characterized so far. Many recent studies deal with the nature of this metastable product. Several energetically competing carbon phases were proposed (W- and M-carbon\cite{OgaJCP2006}, bct C$_4$) as plausible structure solutions, based on estimating transition pressures, goodness of fit of X-ray diffraction data and band gaps\cite{p10,p11,p12}. The intrinsic problematic of stacking faults in the pristine graphite, and Raman evidence of amorphisation suggest a mixture of different phases in the compressed material. The two energetically most preferable candidates so far (M- and W-carbon) can be described (in terms of topology) as corrugated graphene sheets interconnected by an alternating sequence of odd rings (pentagons and heptagons) fused into a 5+7 pattern. This odd-ring topology formally results from connecting puckered graphene layers aligned in a particular way.  On the other hand, further compressing M- or W-carbon can produce different diamonds polytypes. Therefore, a larger variety of intermediate hard structures can in principle be expected.

In this Communication, we further unfold the structural diversity of
sp$^3$-carbon phases. We base our approach on metadynamics simulations
of structural transformations\cite{p13,2006NatMa} and topological enumeration to
efficiently scan configuration space. We report on energetic, mechanical
and electronic properties of four new tetrahedral carbon phases, and insist
on the different underlying graphitic pattern connected with the formation
of a particular topology. We show how distinct topologies with 5+7
(odd-odd), but also 5+8 (odd-even) and 4+6+8 (even-even-even) ring pattern
can do for different mechanical responses.

\begin{figure}
\begin{center}
\includegraphics[width=0.45\textwidth,keepaspectratio]{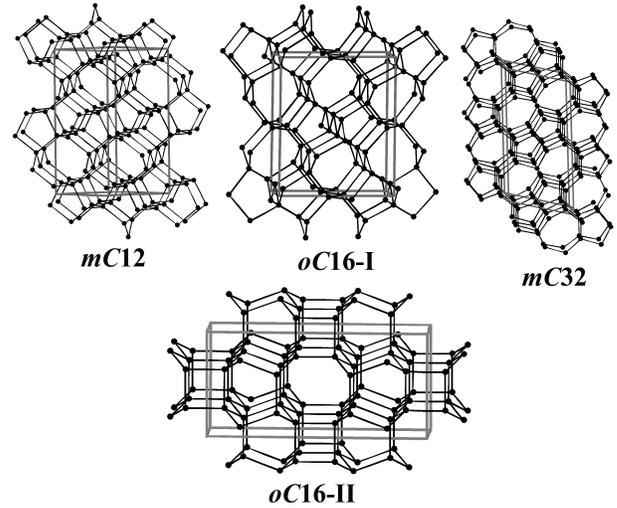}
\end{center}
\caption{\label{Fig_1}Crystal structures of novel carbon phases. oC16-I and mC32 are characterized by 5+7 odd-odd ring pattern. mC12 represents a distinct 5+8 odd-even ring topology, while oC16-II contains even rings only, 4+6+8.}
\end{figure}

\section{methods}
Efficient theoretical approaches to hypothetical carbon modifications,
based e.g. on random techniques, genetic (evolutionary) algorithms, or
accelerated molecular dynamics\cite{b1} result in important discoveries
supporting experiments\cite{p14,p15}. In some approaches the use of graph
theoretical methods\cite{Strong} represents a means of increasing the
sampling efficiency of carbon configuration. It was indeed a graph-theoretical approach that allowed to derive all possible sp$^3$-carbon allotropes with 4 atoms per cell (including the recently rediscovered bct C$_4$)\cite{Strong}. Metadynamics on the other hand explores the energy landscape along collective reaction coordinates, which in case of high pressure polymorphs is represened by the simulation box
itself. While metadynamics does not require prior knowledge of the energy
landscape under investigation, its sampling efficiency improves on
combining many independent runs started from different initial
configurations. Additionally, the number of atoms per simulation box is
critical for capturing a particular atomic configuration. Diamond and
lonsdaleite are important metastable forms of carbon. They can appear in
the same metadynamics run only if the number of atoms in the box is at least 4
and multiple thereof. Similarly, including a minimum of 3 atoms (or
multiple thereof) is sufficient to find a dense carbon with quartz
topology, recently suggested from evolutionary algorithms\cite{PBE_TAB2}.

\begin{table*}
\caption{\label{Table-1}Crystal structure information for novel carbon phases at 0 GPa}
\begin{ruledtabular}
\begin{tabular}{c c c c c c}
  Pearson & Space Group & Wyckoff \\
  Symbol & Cell & position & \raisebox{1.5ex}{x} & \raisebox{1.5ex}{y} & \raisebox{1.5ex}{z} \\
  \hline
  mC12 & C2/c \\
  & a= 3.4242; b=8.5218; & \raisebox{1.5ex}{4e} & \raisebox{1.5ex}{0} & \raisebox{1.5ex}{0.80280} &\raisebox{1.5ex}{3/4} \\
  & c= 3.7012; $\beta$=138.96$^{\circ}$ & \raisebox{1.5ex}{8f} & \raisebox{1.5ex}{0.84662} & \raisebox{1.5ex}{0.91988} &\raisebox{1.5ex}{0.95940} \\
  \hline
  mC32 & C2/m & 8j & 0.46444 & 0.68220 & 0.12680 \\
  & a=9.7242; b=4.2932; & 8j & 0.94998 & 0.68122 & 0.59997 \\
  & c=4.8617; $\beta$=103.96$^\circ$ & 8j & 0.30907 & 0.68472 & 0.43555 \\
  & &  8j & 0.18908 & 0.68609 & 0.87688 \\
  \hline
  oC16-I & C222$_1$ & 4a & 0.43209 & 1/2 & 0 \\
  & a=6.6698; b=5.5609; c=2.5119 & 4b & 1/2 & 0.08196 & 1/4 \\
  & & 8c & 0.81701 & 0.76297 & 0.11960 \\
  \hline
  oC16-II & Cmmm & 8p & 0.66672 & 0.68505 & 0.0 \\
          &a=8.8134; b=4.2743; c=2.5281 & 8q & 0.58903 & 0.81586 & 1/2 \\
\end{tabular}
\end{ruledtabular}
\end{table*}

To systematically include known and find novel carbon forms, metadynamics
runs were performed on simulation boxes comprising three, four, six, eight,
twelve and sixteen carbon atoms, respectively. A similar approach has been
shown to work well in connection with plain MD to search for ice phases
\cite{Buch2005,Buch2006}. Quasi-random four-connected nets were used as
starting configurations. We note here that the application of metadynamics
in this case is slightly different from its typical use for simulation of
crystal-crystal structural phase transitions, as recently reviewed in
Refs.\cite{csp_wiley2010,epjb2011}. While in both cases the simulation cell
is used as order parameter, here one instead starts from a disordered
configuration in a small cell and searches for low-energy configurations
representing crystalline structures with a given number of atoms in the unit
cell. Each run was typically consisted of about 25000 metasteps.
Within each metastep MD was performed in the NVT ensemble for at least 0.5 ps at 300 K.
In these preliminary scans the tight binding Tersoff
potential\cite{p16} was used, which ensured rapid and reliable structure
evolution thanks to its good description of sp$^2$/sp$^3$ carbons.
Molecular dynamics in the NVT-ensemble was performed with the CP2K
code\cite{p17,p18}. Structure diversity was judged by calculating vertex
symbols, which contain information on all shortest rings meeting at each
atom, and coordination sequences, as implemented in the TOPOS
package\cite{TOPOS}. Both topological descriptors are widely used e.g. for
the topological characterization of zeolites\cite{TOPOS}.  Metadynamics
trajectories contained many foam-like structures with mixed sp$^2$-sp$^3$
carbon atoms and a few sp$^3$-allotropes. In case of new tetrahedral
structure types, inferred from their topology, ideal space groups and
asymmetric units were identified with the Gavrog Systre
package\cite{SYSTRE}. In a subsequent set of runs, candidate structures
were studied with respect to their transformability into diamond by
metadynamics simulations using SIESTA\cite{p19,p20} as DFT/MD layer.

In the initial metadynamics runs the choice of large pressure values is less critical. It is rather the number of atoms in the simulation box that decides whether a particular topology can be visited at all within a single metadynamics run. The structures presented in the following were harvested from metadynamics runs performed at 0 and 5 GPa, with 6, 8 and 16 atoms in the simulations box.

On idealized structures variable cell conjugate-gradient relaxation was performed within density functional theory (GGA, PBE) as implemented in the SIESTA package\cite{p19,p20}. Electronic states were expanded by a double-ζ basis set with polarization functions (DZP). Core states (1s$^2$) were described by norm-conserving Troullier–Martins pseudopotentials \cite{p21}. The charge density was represented on a real-space grid with an energy cutoff of 200 Ry. Forces were relaxed to less than 0.01 eV/{\AA}. Convergence with respect to the number of k-points was carefully checked. 

\section{results and discussion} Small boxes of 2 and 3 atoms expectedly produced cubic diamond and quartz, respectively. With four atoms both cubic and hexagonal diamond (Lonsdaleite) were collected. From 6, 8 and 16 atoms metadynamics three novel structures were found, two monoclinic (mC12 and mC32, Fig.~\ref{Fig_1}) and one orthorhombic (oC16-I, Fig.~\ref{Fig_1}). From further propagating oC16-I in metadynamics runs at 100 GPa, oC16-II (Cmmm) was found. Their symmetries and structural parameters are summarized in Table~\ref{Table-1}. All phases correspond to a stacking of corrugated graphene layers interconnected by alternating sequence of pentagons and heptagons (oC16-I and mC32, Fig.~\ref{Fig_1}), like for M- and W-carbon. Alternatively, pentagons and octagons, or square and octagons can also be placed between puckered graphitic layers like it is realized in mC12 and oC16-II, respectively~(Fig.~\ref{Fig_1}). Physical properties of the new allotropes are compared with those of known structures in Table~\ref{Table-2}. In terms of volume per atom, oC16-I is the densest, hardest structure, closely followed by oC16-II and mC12. With a calculated band gap of 4.5 eV, oC16-I is also the structure closest to diamond. 

\begin{figure}
\begin{center}
\includegraphics[width=0.45\textwidth,keepaspectratio]{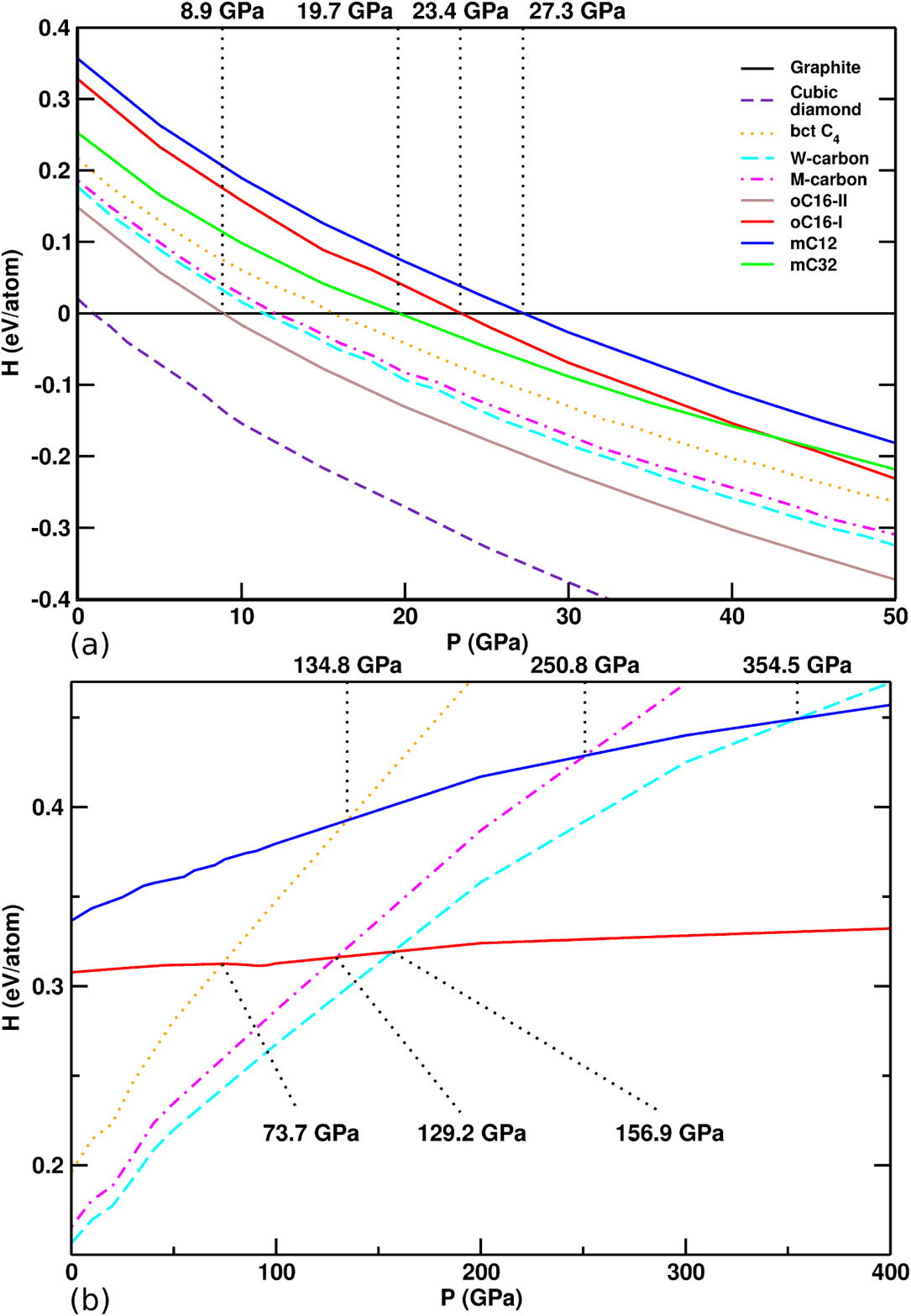}
\end{center}
\caption{\label{Fig_2}(a) Enthalpies (relative to graphite) of different carbon allotropes; (b) enthalpies of certain sp$^3$-allotropes (relative to diamond) in the high pressure range. The colours are the same as in (a).}
\end{figure}

The stability of different carbon phases in a wide pressure range is presented in Fig.~\ref{Fig_2}. At elevated pressures, the novel allotropes become more stable than graphite, Fig.~\ref{Fig_2}b. The transition pressures are similar for the mC32 and oC16-I structures (19.7 and 23.4 GPa, respectively) and much higher (by ~10 GPa) for mC12. mC12 and particularly oC16-I are stabilized upon increasing pressure. Furthermore, the stability of oC16-I remains basically constant (up to 400 GPa) whereas M- and W-carbon rapidly become energetically unfavourable above 100 GPa.

Bulk moduli (B$_0$) were obtained by fitting total energy as a function of volume to the third-order Birch-Murnaghan equation of state (Table~\ref{Table-2}). Strickingly, oC16-I is harder than M- and W-carbon, although less stable below 129.2 GPa (Fig.~\ref{Fig_2}). On the contrary, oC16-II features a lower enthalpy, its gap is nonetheless intermediate between lonsdaleite and bct C$_4$, which is structurally also the case. By inspection of~Fig.~\ref{Fig_1}, motifs of lonsdaleite and bct C$_4$ can easily be recognized.

\begin{figure}
\begin{center}
\includegraphics[width=0.45\textwidth,keepaspectratio]{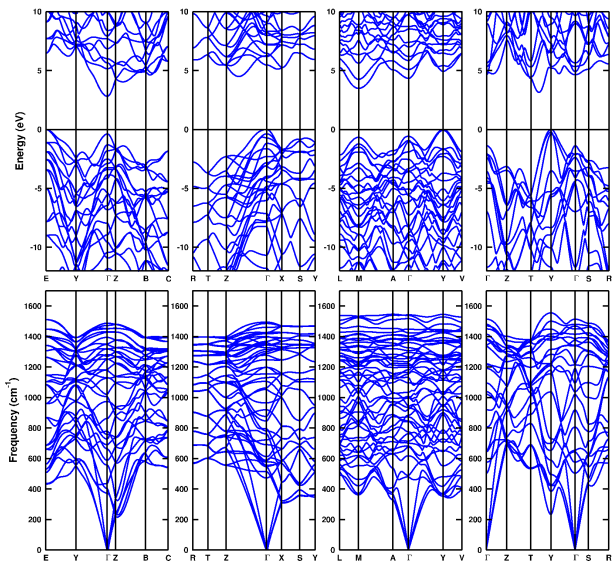}
\end{center}
\caption{\label{Fig_3} Electronic band structures (top) and phonon dispersion curves (bottom) for (from left to right) mC12, oC16-I, mC32 and oC16-II carbon phases.}
\end{figure}

The electronic band structures of mC12, oC16 (I and II) and mC32 at 50 GPa are shown in~Fig.~\ref{Fig_3}. The structures are insulating with indirect band gaps in the range 2.8-4.5 eV. The gaps do not depend on the pressure up to 50 GPa. All the gaps are smaller than in diamond, but similar to those of M- and bct-carbon.

\begin{table*}
\caption{\label{Table-2}Calculated equilibrium volume (V$_0$), bulk modulus (B$_0$) and band gaps (E$_g$). All values refer to zero pressure.}
\begin{ruledtabular}
\begin{tabular}{c c c c c c}
Structure & Method & V$_0$ ({\AA}$^3$)& B$_0$ (GPa) & E$_g$(eV) & H(Gpa)\footnotemark[1]\\
\hline
& This work &5.79 & 424.2 & 4.19 & 87.3\\
\raisebox{1.5ex}{Diamond} & PBE\cite{PBE_TAB2} & 5.70 & 431.1 & 4.2 \\
\hline
mC12 & This work & 5.91 & 399.5 & 2.82 & 84.4 \\
\hline
oC16-I & This work & 5.82 & 411.0 & 4.5 & 85.8 \\
\hline
mC32 & This work & 6.16 & 384.5 & 3.47 & 70.2 \\
\hline
oC16-II & This work & 5.95 & 408.4 & 3.15 & 84.4 \\
\hline
W-carbon & This work & 6.04 & 391.8 & 4.35 & 83.1 \\
\hline
& This work & 6.06 & 392.6 & 3.51 & 82.7 \\
\raisebox{1.5ex}{M-carbon} & PBE\cite{PBE_TAB2} & 5.97 & 392.7 & 3.60 \\
\hline
& This work & 6.11 & 393.4 & 2.6 & 82.0 \\
\raisebox{1.5ex}{Bct C$_4$} & PBE\cite{PBE_TAB2} & 6.01 & 411.4 & 2.7 \\
\footnotetext[1]{According to the method of A.O. Lyakhov and A.R. Oganov\cite{p22}}
\end{tabular}
\end{ruledtabular}
\end{table*}

Phonon dispersion curves were calculated within a pressure range up to 100 GPa. No imaginary frequencies were observed throughout the whole Brillouin zone, confirming the dynamical stability of the intermediate sp$^3$ structures (Fig.~\ref{Fig_3}). Isothermic-isobaric molecular dynamics simulations (300 K, 1 atm, 3 ps) also confirmed the stability of the new phases.  

\begin{figure}
\begin{center}
\includegraphics[width=0.45\textwidth,keepaspectratio]{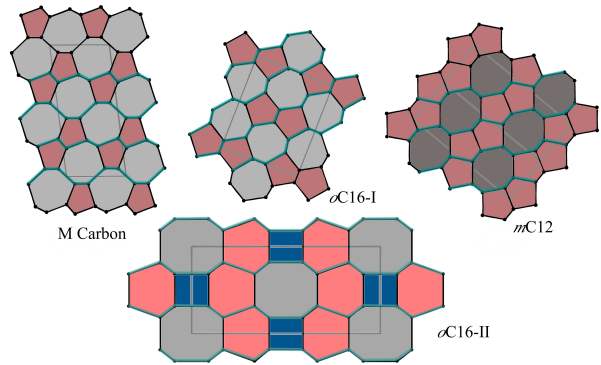}
\end{center}
\caption{\label{Fig_4} Comparison of M-carbon with oC16-I, mC12 and oC16-II with respect to their underlying puckered graphitic stacking. Layers are highlighted in turquoise.}
\end{figure}

Fig.~\ref{Fig_4} shows the relation between the discovered
structures and the graphene layer stackings they are deriverd from. This
information can be obtained by deconstructing the structures and
looking for graphitic layers within the lattices. For the novel structures,
the matching we are presenting is supported by the metadynamics runs, where
a (fully or partially) graphitic structure is a typicall precursor of the
sp$^3$ phases, along the simulation time coordinate.

\begin{figure}
\begin{center}
\includegraphics[width=0.45\textwidth,keepaspectratio]{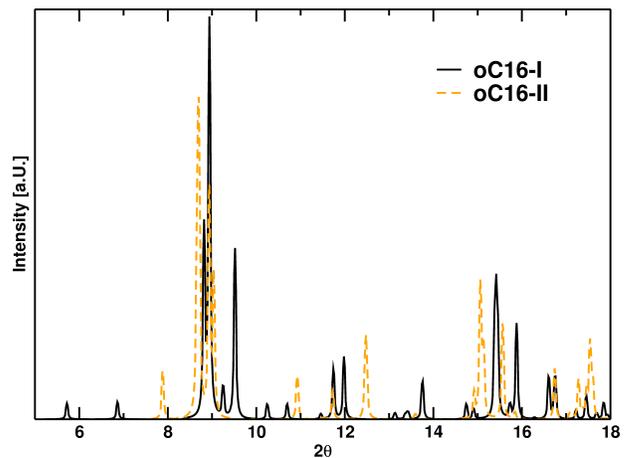}
\end{center}
\caption{\label{Fig_5} Simulated XRD pattern for oC16-I and oC16-II carbon phases ($\lambda=0.3329${\AA}). The structural data are those of Table~\ref{Table-1}.}
\end{figure}

In general, we notice that hardness and band gap are diversly distributed
among the phases. In the effort of providing an answer to the outstanding
question of hard and transparent sp$^3$ carbon, oC16-II and oC16-I appear
as better candidates as hitherto suggested, the former particularly for its
stability, for a really transparent band gap and hardness the latter.  In Fig.~\ref{Fig_5} we present the simulated XRD patterns of oC16-I and oC16-II. With reference to the experimental pattern\cite{p8}, the relevant regions between 8.5-10$^\circ$ as well as 14.5-17$^\circ$ are similarly populated. Intermediate peaks between 10-14$^\circ$ can better distinguish between the two structures,  are however depleted in the experiments\cite{p8} such that the experimental match is substantially the same for oC16-I and oC16-II.  

Since superhard graphite is not synthetized from the gas phase, which would
probably produce oC16-II as the only product due to its lowest enthalpy,
in the real experiment much will depend on the nature of the starting graphitic material, and on
the particular nucleation history, which would favor one pattern in the
phase growth phase. In this context, the overall stability of a particular structure
is not the only parameter. The importance of this point of view has
been recently pointed out~\cite{Parri_nmat}, and dedicated investigations are ongoing.

In conclusion, we have presented four novel sp$^3$ carbon materials,
derived from combining metadynamics and topology to achieve higher scan
efficiency. Two structures, oC16-I and oC16-II stand out for hardness and
band gaps, and should be considered in assessing the nature of the product
of graphite cold compression.

\begin{acknowledgments}
  R.M. was supported by the Slovak Research and Development Agency
  under Contract No. APVV-0558-10 and by the project implementation
  26240120012 within the Research \& Development Operational Programme
  funded by the ERDF. S.L. thanks the DFG for support under the priority project SPP 1415, as well as ZIH Dresden for the allocation of computational resources. Finally, we thank Prof. A.R. Oganov for inspiring discussions.
\end{acknowledgments}

\bibliography{carbo.bib}


\end{document}